%% file: PTSD_in_the_wild.tex
\documentclass[journal]{IEEEtran}

\usepackage{hyperref}

\ifCLASSINFOpdf
\else
\fi
\usepackage{graphicx}

\usepackage{amsfonts,amssymb,amsmath}
\usepackage{xcolor}

\usepackage{caption}
\usepackage{subcaption}

\usepackage{tikz}
\usepackage{mdframed}
\usepackage[export]{adjustbox}
\usepackage{dashrule}
\usepackage{multirow}
\usepackage{makecell}
\usepackage{pgf-pie}
\usepackage{pgfplots}

\usepackage{array}
\usepackage{longtable}
\usepackage{tikz}
\usetikzlibrary{positioning}

\usepackage{array}
\newcolumntype{C}[1]{>{\centering\let\newline\\\arraybackslash\hspace{0pt}}m{#1}}

\begin{document}
%
\title{PTSD in the Wild: A Video Database for Studying
Post-Traumatic Stress Disorder Recognition in Unconstrained Environments}
%
%
%

\author{Moctar Abdoul Latif Sawadogo,
        Furkan Pala,
        Gurkirat Singh,
        Imen Selmi,
        Pauline Puteaux,
        and~Alice~Othmani
        \thanks{M.A.L. Sawadogo is with LISSI laboratory of University Paris-Est Créteil and University Paris City in France. E-mail: moctar-abdoul-latif.sawadogo@u-pec.fr \\
        F. Pala is with LISSI laboratory of University Paris-Est Créteil in France and Istanbul Technical University (ITU) in Turkey. E-mail: pala18@itu.edu.tr \\
        G. Singh is with LISSI laboratory of University Paris-Est Créteil in France and the National Institute of Technology Warangal in India. E-mail: gs832025@student.nitw.ac.in \\
        I. Selmi is with LISSI laboratory of University Paris-Est Créteil in France and The National Engineering School of Sfax (ENIS) in Tunisia. E-mail: imen.selmi@enis.tn \\
        P. Puteaux is with the French National Centre for
         Scientific Research (CNRS) in France. E-mail: pauline.puteaux@cnrs.fr \\
        A. Othmani is with LISSI laboratory of University Paris-Est Créteil. E-mail: alice.othmani@u-pec.fr 
        }
\thanks{This research work is realized in the Laboratoire Images, Signaux et Systèmes Intelligents (LiSSi)- EA-395, Université Paris-Est Créteil (UPEC), \\
Corresponding author: Dr. Alice Othmani, E-mail: alice.othmani@u-pec.fr}
}

\maketitle

\begin{abstract}

POST-traumatic stress disorder (PTSD) is a chronic and debilitating mental condition that is developed in response to catastrophic life events, such as military combat, sexual assault, and natural disasters. 
PTSD is characterized by flashbacks of past traumatic events, intrusive thoughts, nightmares, hypervigilance, and sleep disturbance, all of which affect a person's life and lead to considerable social, occupational, and interpersonal dysfunction.
The diagnosis of PTSD is done by medical professionals using self-assessment questionnaire of PTSD symptoms as defined in the Diagnostic and Statistical Manual of Mental Disorders (DSM). In this paper, and for the first time, we collected, annotated, and prepared for public distribution a new video database for automatic PTSD diagnosis, called PTSD in the wild dataset. The database exhibits "natural" and big variability in acquisition conditions with different pose, facial expression, lighting, focus, resolution, age, gender, race, occlusions and background.
In addition to describing the details of the dataset collection, we provide a benchmark for evaluating computer vision and machine learning based approaches on PTSD in the wild dataset. In addition, we propose and we evaluate a deep learning based approach for PTSD detection in respect to the given benchmark. The proposed approach shows very promising results. Interested researcher can download a copy of PTSD-in-the wild dataset from: 
\url{http://www.lissi.fr/PTSD-Dataset/}

\end{abstract}


%
\IEEEpeerreviewmaketitle

\section{Introduction}
%
%
%
%
\IEEEPARstart{P}{ost}-traumatic stress disorder (PTSD) is a chronic mental condition resulting after exposure to a traumatic event. These events generally involve direct threat, or represent a real risk of death or serious injury like natural disasters, sexual assault, military combat experience or even during a major stressful life experience like divorce or unemployment. The symptoms of PTSD are divided into three symptom clusters: reexperiencing, avoidance, and hyperarousal. In addition, trauma survivors often experience guilt, dissociation, alterations in personality, difficulty with affect regulation, and marked impairment in ability for intimacy and attachment. Disorders comorbid with PTSD include depression, substance abuse, other anxiety disorders, and a range of physical complaints.
Over the past several decades, considerable progress has been made in the development and empirical evaluation of assessment instruments for measuring trauma exposure and PTSD as well as related syndromes, such as acute stress disorder. The measures that have been developed, including self-report questionnaires, structured interviews~\cite{De_Beurs2020-nr}, and psychophysiological procedures~\cite{Bauer2013-hu}, have been extensively validated and many have been widely adopted internationally.

Traditionally like most other debilitating psychological disorders, it was diagnosed by medical professionals in a clinical setting. Diagnosing the ailment would involve a comprehensive psychological evaluation comprising questionnaires and multiple in-person interview sessions. Occasionally blood tests are an additional requirement for filtering out the specific psychological affection. Collecting information through a self-report has limitations. People are often biased when they report on their own experiences. And these self-reports are subject to several biases and limitations, such as the introspective ability: the subjects may not be able to assess themselves accurately or the rating scales bias : Rating something yes or no can be too restrictive, but numerical scales also can be inexact and subject to individual inclination to give an extreme or middle response to all questions.

Early and automatic detection of PTSD can help reduce its effect. Hence, the interest in early detection and the audio-visual mechanisms involved in post-traumatic stress detection. To the best of our knowledge, no public video dataset exists for PTSD recognition and prediction. To meet this need, We present in this paper, a new publicaly available dataset for academics to facilitate the development of new approaches and tools for the automatic diagnosis of post traumatic stress disorder. This dataset as well as the proposed deep learning based approach for PTSD recognition will serve as a valuable starting point for scholars interested in affective computing of mental disorders and audiovisual data to investigate and to study PTSD.

The rest of this paper is organized as follows. In Section~\ref{sec:sota}, we detail current state-of-the-art methods related to PTSD diagnosis from clinical videos and existing video datasets. Section~\ref{sec:dataset} describes the new video dataset we constructed, named PTSD-in-the-wild dataset.
In Section~\ref{sec:baseline}, we give explanations on the used baseline models (audio, visual and text). Section~\ref{sec:results} presents experimental results and finally, this paper is concluded in Section~\ref{sec:conclusion}.

\section{Related works and databases}
\label{sec:sota}

\begin{table*}[]
\centering
\begin{tabular}{|l|l |l | l|}
\hline
\textbf{Ref.} & \textbf{Dataset}  & \textbf{Modalities}  & \textbf{Public}\\
\hline
 \cite{islam2018transfer} &  FEMH (Far Eastern Memorial Hospital) Dataset & Audio &  NO \\ \hline
\cite{gratch2014distress} & eDIAC-WOZ  & Audio + visual features & only for academic research under request \\ \hline
  \cite{mclean2020aurora}  & Aurora & Audio + text  & NO \\ \hline
  \cite{stappen2021muse}  & Ulm-TSST (Muse challenge) & Video + ECG + EDA + Respiration + Heart Rate &  only for academic research under request\\ \hline
\end{tabular}
\caption{Existing audio and video datasets for stress and PTSD diagnosis.}
\label{table:datasets}
\end{table*}

In this section, we present related works and databases for PTSD and stress diagnosis. In Section~\ref{subsec:methods}, we first detail current state-of-the-art PTSD diagnosis from clinical videos methods. In Section~\ref{subsec:datasets}, we describe existing video datasets, which were all acquired in lab-controlled conditions.

\subsection{PTSD diagnosis from clinical videos}
\label{subsec:methods}

 In recent years, Artificial Intelligence (AI)-based systems have become immensely popular for psychological diagnosis due to their scalability, flexibility and convenience. A system for the mental screening of military personnel -- where audio interviews were fed into the \textit{Sensibility technology ST Emotion} framework for emotion recognition -- has been proposed in~\cite{tokuno2011usage}. Soldiers who spent more time on a particular mission had higher stress levels and lower happiness compared to those who served for a shorter time.

In \cite{schultebraucks2022deep}, audio-visual features comprising of face, voice, speech content and movement were extracted from medical interview recordings and fed into a neural network for PTSD and MDD (Major Depressive Disorder) prediction. It was found that textual features contributed the most to an accurate prediction. In \cite{banerjee2019deep}, only the audio data was used. To compensate for the small sized PTSD dataset that was available, they decided to use transfer learning. A deep belief network was initially trained on audio data to learn general voice features after which fine-tuning was performed on the PTSD dataset.

The accuracy of audio-based PTSD classifiers could be increased by using multi-view learning, as shown in~\cite{zhuang2014improving}. The classifier itself was trained on EEG (electroencephalography) and audio data, but used only audio signals for prediction during inference. A prediction on audio data is also performed in~\cite{gupta2022toxgb}. After preprocessing, prosodic, excitation and vocal tract features were extracted and fed to an architecture based on extreme gradient boosting with teamwork-based optimization for prediction. In \cite{rozgic2014multi}, it was shown that elicitation of multimodal neurophysiological responses to audio-visual stimuli was a suitable alternative to using clinical interviews as input for PTSD diagnosis. Neurophysiological responses comprising of ECG (electrocardiography), EEG, GSR (galvanic skin response), head motion, video, speech, along with the emotions evoked were fed into an SVM classifier. A comprehensive study on post-traumatic neuropsychiatric sequelae is proposed in~\cite{mclean2020aurora}. It provided a huge multimodal dataset on which PTSD models could be trained.

Owing to the overlap between PTSD and the symptoms of other mental disorders, it is important to go over the work done on the latter as well. In~\cite{yang2017hybrid}, audio-visual data was used along with the text. The model utilized a DCNN and DNN based architecture to get predictions from audio-video modalities, while paragraph vector, SVM and random forest were utilized to predict from the textual input. Fusion using a multivariate regression model fused the data from all three modalities for the final prediction.


\subsection{Existing video datasets}
\label{subsec:datasets}
Few video datasets for PTSD and stress diagnosis exist, as shown in Table~\ref{table:datasets}. There are only three datasets with audio data for PTSD recognition: eDAIC-WOZ~\cite{gratch2014distress}, FEMH (Far Eastern Memorial Hospital)~\cite{islam2018transfer} and Aurora~\cite{mclean2020aurora}. Moreover, these datasets were all acquired in lab-controlled conditions.

\paragraph{eDAIC-WOZ~\cite{gratch2014distress}}
It is an extended version of DAIC-WOZ dataset for detection of PTSD from an interview without any human intervention. This is an audio visual dataset that was also used for the AVEC 2019 Challenge. The dataset contains clinical interviews which were collected as part of a larger effort to create a computer agent which identifies verbal and non verbal signs of mental illness. It contains 189 interviews ranging between 7-33 minutes. This dataset is made publicly available only for academics researchers under request and signature of an End User License Agreement (EULA).

\paragraph{FEMH dataset (Far Eastern Memorial Hospital)~\cite{islam2018transfer}} 
It consists of voice signals from 200 patients labeled as normal (50), neoplasm (40), phonotrauma (60) and vocal palsy (50). These voice signals contain 3-second sustained /a:/ sounds, with a
sampling rate of 44.1KHz and a 16-bit resolution. The voice recording lengths are between 5 seconds to 40 seconds.

\paragraph{Aurora dataset~\cite{mclean2020aurora}} RecOvery afteR traumA (AURORA) dataset conducts a large scale (n = 5,000 target sample) in-depth assessment of adverse post-traumatic neuropsychiatric sequelae (APNS) development. These APNS include traditionally categorized outcomes such as PTSD, depression, MTBI (minor traumatic brain injury), and regional or widespread pain. The 5,000 patients were screened, recruited and receive initial baseline evaluation, including blood collection and psychophysical, survey and neurocognitive evaluation. They were closely monitored over the next 8 weeks using a wrist wearable, a smart phone app and daily flash surveys weekly, web-based neurocognitive tests, periodic mixed-mode surveys, serial saliva collection, deep phenotyping blood collection, fMRI (functional magnetic resonance imaging), psychophysical evaluation and then followed less intensively using similar procedures including deep phenotyping over the remainder of a 52-week follow-up period.

Two others video datasets exist for recognizing stress and positive/negatives emotions. We notice that these two datasets could be interesting for transfer knowledge from stress and negative emotions domain to stress post-traumatic domain. These two datasets are : Ulm-TSST and MuSe-CaR datasets.

\paragraph{Ulm-TSST~\cite{stappen2021muse}} 
This dataset contains about 10 hours long audio visual, as well as biological recordings such as ECG, EDA (electrodermal activity), respiration and heart rate. It consists of a richly annotated dataset of self reported and external dimensional ratings of emotion and mental well being. It has videos of 105 participants out of which 69.5\% are female which are aged between 18 to 39 years. This dataset is available online for researchers who fulfil the requirements of the EULA.

\paragraph{MuSe-CaR~\cite{DBLP:journals/corr/abs-2101-06053}}
It is a large multimodal dataset which contains about 303 videos, \textit{i.e.} over 40 hours of user-generated video material. This dataset mainly focuses on how positive and negative sentiments as well as emotional arousal are linked to the person. This dataset is extensively annotated and is available online for researchers upon request.

\section{PTSD in-the-wild dataset collection}
\label{sec:dataset}

In this section, we describe a new video dataset called PTSD in-the-wild dataset for studying PTSD in unconstrained environments. In Section~\ref{subsec:videos_collection}, we explain the procedure of videos collection from the web. In Section~\ref{subsec:annotation}, the videos annotation process is detailed and the categories of trauma and their distribution are presented.

\begin{figure*}[!ht]
    \centering
\subfloat[][\label{im:ex1_ptsd}]{\includegraphics[height=3.5cm]{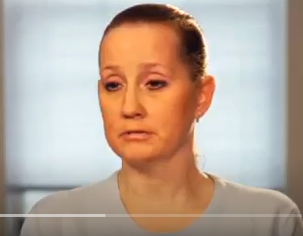}}\hspace{1mm}
\subfloat[][\label{im:ex2_ptsd}]{\includegraphics[height=3.5cm]{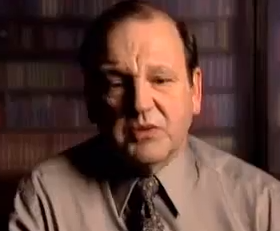}}\hspace{1mm}
\subfloat[][\label{im:ex1_not_ptsd}]{\includegraphics[height=3.5cm]{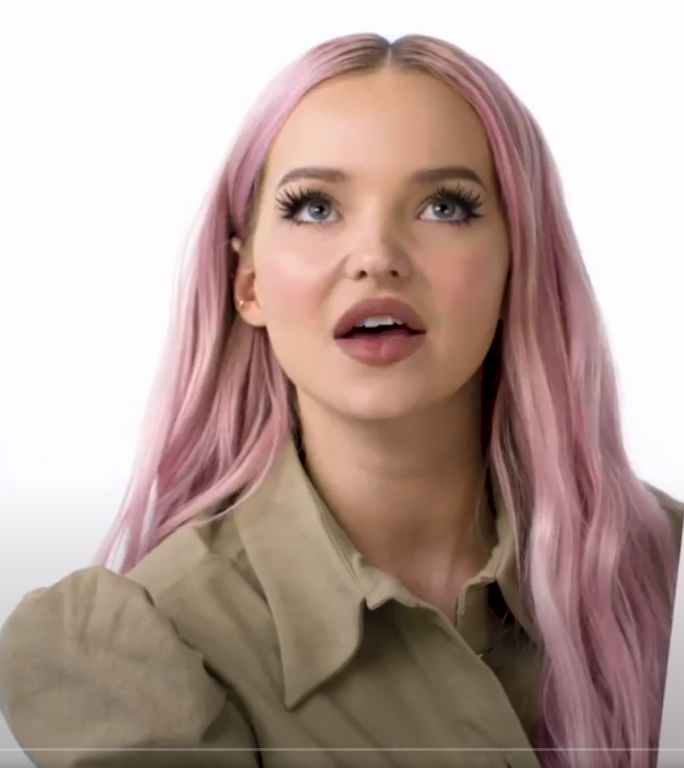}}\hspace{1mm}
\subfloat[][\label{im:ex2_not_ptsd}]{\includegraphics[height=3.5cm]{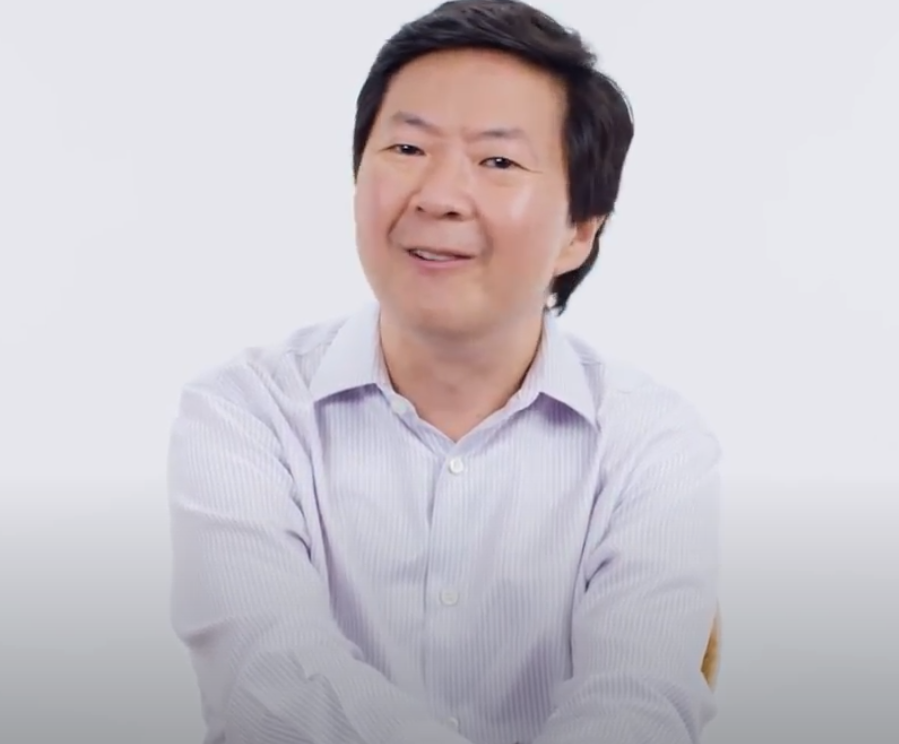}}\hspace{1mm}
    \caption{\label{fig:examples} Four videos frames from PTSD-in-the-wild dataset of two subjects with PTSD ((a) and (b)) and two subjects without PTSD ((c) and (d)). (a) A woman who had PTSD after war. She was on a mission and suddenly a bomb exploded. She found out that she has PTSD in October 2004. (b) Man who suffered from PTSD after a plane crash. He was the captain of the plane. After the plane crash, his body was damaged a lot and he always felt guilty. (c) Dove Cameron answers the Web’s Most Searched Questions with WIRED auto-complete interview. Dove Olivia Cameron is an American singer and actress. She played a dual role as the eponymous characters in the Disney Channel comedy series Liv and Maddie. (d) Ken Jeong answers the Web’s Most Searched Questions with WIRED auto-complete interview. Kendrick Kang-Joh Jeong is an American stand-up comedian, actor, producer, writer and licensed physician.
    }
\end{figure*}

\subsection{Videos collection from the web}
\label{subsec:videos_collection}
The procedure of collecting this dataset was realized semi-automatically. It was done first by downloading more than 1,000 videos from YouTube and then, by choosing the videos of interviews to predict whether a person has a PTSD or not.

In our search, we used keywords such as: "PTSD", "Understanding PTSD", "my story with PTSD", "celebrities Answers the Web's Most Searched Questions", "interviews with celebrities", \textit{etc}. For data with PTSD, we selected a group of people who had experienced a traumatic event like an accident or a terrorist attack.
We collected this data from news, entertainment and medical channels on YouTube like Make the connection\footnote{https://www.youtube.com/c/VeteransMTC }, Veterans Health Administration\footnote{https:
//www.youtube.com/c/VeteransHealthAdmin} and GQ\footnote{GQ: https://www.youtube.com/c/GQ}.
For data without PTSD, we selected a group of celebrities and artists who have been interviewed.

All these videos are in English. Moreover, there are different age groups and genders in this data, as well as nationalities.
The average video duration is about 2 minutes. In most of the videos, the interviewee speaks more than the interviewer. There are also videos without any interviewer: only an interviewee talks and answers questions.

In Fig.~\ref{fig:examples}, as examples, we present 4 pictures extracted from videos of our dataset. Pictures depicted in Fig.~\ref{fig:examples}.a and Fig.~\ref{fig:examples}.b are taken from videos of people with PTSD, while those in Fig.~\ref{fig:examples}.c and Fig.~\ref{fig:examples}.d show people without PTSD. One can see from Fig.\ref{Fig:gender_distribution} that the two genders are represented both in the With PTSD and Without PTSD classes. In Fig.~\ref{fig:examples}, the interviewees with PTSD seem to be troubled, whereas it is not the case of the interviewees that do not have PTSD. Note also that, within the 'With PTSD' class examples, the PTSD disorder originates from two different types of trauma: war veteran (Fig.~\ref{fig:examples}.a) and plane crash (Fig.~\ref{fig:examples}.b).

\subsection{Annotation}
\label{subsec:annotation}
The videos were annotated by human annotators. Four people participated to the annotation of the dataset. After extraction, each video was watched by an annotator and different labels were given. Apart from the class label, other categories of information were collected. These are the class, the gender and the type of developed trauma. In the paragraphs below, we explain how the dataset was annotated and how the categories were determined.

\subsubsection{Class label} 
Videos were classified into 'With PSTD' and 'Without PTSD'. Videos labelled as 'With PTSD' are about people who developed PTSD after a traumatic event and they are talking about their trauma during an interview. Videos labelled as Without PTSD are interview videos about random topics which are not PTSD-related. The videos are labelled by taking into account the speech of the interviewee and the description of the video. If the interviewee mentions that he was diagnosed with PTSD or it is mentionned in the video's description that the interviewee has PTSD, then the video is labelled as 'With PTSD'. On the other hand, if neither the speech nor the description is about PTSD or a trauma, the video is labelled as 'Without PTSD'. The dataset contains 634 videos equally balanced: there are 317 videos of subjects with PTSD and 317 videos of healthy control subjects with no PTSD symptoms.



\subsubsection{Gender} 
\begin{figure*}[]

    \begin{subfigure}[b]{0.45\textwidth}
 
      \scalebox{0.6}{\input{./graphics/trauma_distribution}}
      \caption{Distribution of the types of trauma.}
      \label{Fig:trauma_types}
    \end{subfigure}
    \hspace*{\fill} 
    \begin{subfigure}[b]{0.45\textwidth}
      \scalebox{0.85}{\input{./graphics/gender_trauma_distribution}}
      \caption{Distribution of the gender in the dataset. Note
that, in some videos With PTSD, there are more than one
interviewee. This is why the number of interviewees exceeds
the number of videos.}
      \label{Fig:gender_distribution}
    \end{subfigure}
\caption{Distribution of the type of trauma and the gender across the PTSD-in-the-wild dataset. (a) Distribution of the types of trauma. (b) Distribution of the gender in the dataset.}    
\label{Fig:dataset_distribution}
\end{figure*}
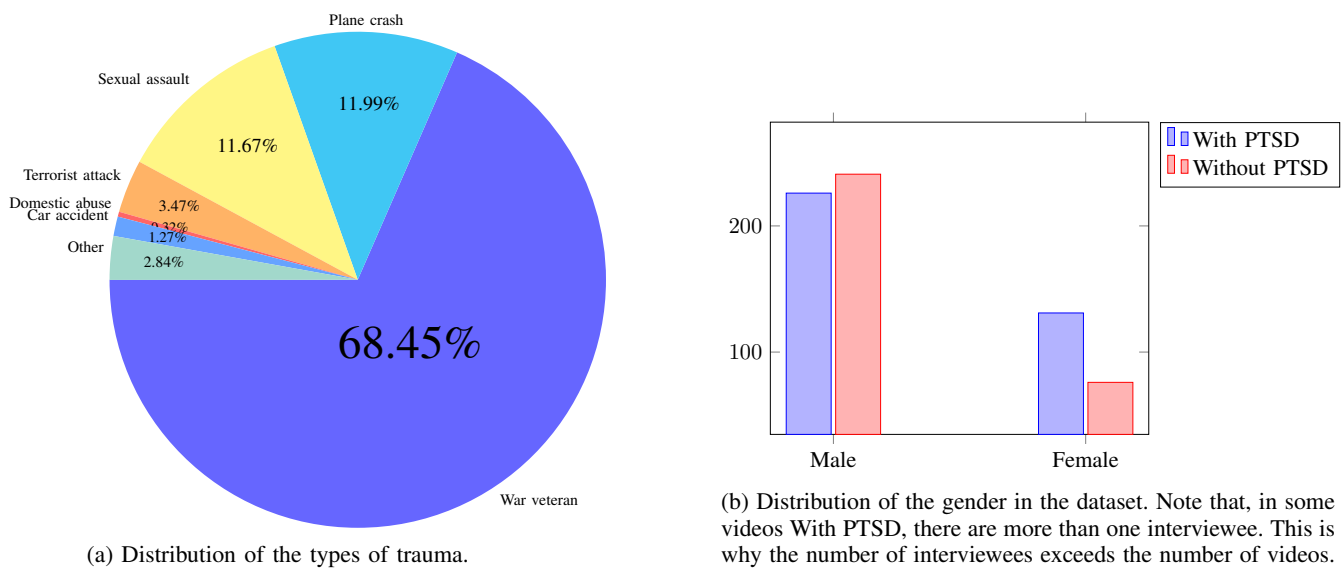

The gender of each interviewee was determined from the video. The considered gender types were Male (M) and Female (F). Some videos have more than one interviewee. Therefore, the number of interviewees exceeds the number of videos. The dataset globally contains 467 male interviewees and 207 female interviewees. There are different gender distributions across the classes With PTSD and Without PTSD. Fig.~\ref{Fig:gender_distribution} summarizes the distribution of the gender within the two labels.


\subsubsection{Type of trauma} 

  The types of trauma were also considered in the annotation of the dataset. A trauma is a response to a terrible event. The type of the trauma concerns the event that caused the trauma. It was determined from the interviewee's speech or from the video's description. Many types were found. They are grouped into the following categories: War veteran, Plane crash, Sexual assault, Terrorist attack, Domestic abuse, Car accident and Other. Fig.~\ref{Fig:trauma_types} shows the count of the different types of trauma in the dataset.
  
  \begin{itemize}
      \item \textbf{War veteran:} This category contains videos of people who have been in the army and have been deployed in hostile territories or war zones. They subsequently developed PTSD at some point during their service or after ending their service. This is the most represented category.
      
      \item \textbf{Sexual assault:} This category contains videos of people who have endured sexual assaults and harassment. Some of the videos in this category are Military Sexual Trauma (MST), they are people in the military who have suffered sexual assaults from other soldiers.
  
      \item \textbf{Terrorist attack:} This category gathers the videos of people who have been direct victims of a terrorist attack or have been witnesses of a terrorist attack. Terrorist attacks are events where violence is used to create fear in a population. Most of the time, it involves killing people. In the collection of the dataset, we considered many such events among which school shootings and September 11 attacks.
  
      \item \textbf{Plane crash:} This videos regroup people who have survived a plane crash or a plane related incident.
      
      \item \textbf{Domestic abuse:} This category contains videos of people who have developed a trauma from domestic abuse by their partner. 
      
      \item \textbf{Car accident:} This category contains videos of people who have been victims or witnessed a terrible car accident which led them to develop PTSD.
      
      \item \textbf{Other:} This category regroups videos of people whose type of trauma could not be classified in any of the aforementioned categories. This contains categories like physical abuse, gang violence and videos of people who have suffered several and repeated traumatic events. Furthermore, people whose type of trauma could not be determined are classified in this category.
 \end{itemize}


\section{Baseline}
\label{sec:baseline}
In this section, we present our baseline for PTSD diagnosis in unconstrained environments, using our proposed PTSD in-the-wild dataset. In Section~\ref{subsec:splits}, we explain how we built sets for training, validation and test. Then, we describe the models we proposed, based on different modalities: audio (Section~\ref{subsec:audio_model}), visual (Section~\ref{subsec:visual_model}) and text (Section~\ref{subsec:text_model}).

\subsection{Training, validation and testing sets\label{subsec:splits}}

Our dataset consists of $317$ videos of subjects with PTSD and $317$ videos of healthy control subjects with no PTSD symptoms.
Two benchmarks are proposed for the evaluation of new approaches for PTSD diagnosis and recognition. Then, for splitting the dataset into training, validation and testing sets, we used the two strategies or partitions:

\paragraph{Train/Validation/Test split} We randomly split the dataset into training, validation and testing sets with the percentages of 80\%, 10\% and 10\%, respectively. Distribution of the two classes of With and Without PTSD in the three sets of training, validation and test can be seen in the Table~\ref{table:single_split_dist}: among the $317$ samples for each class, we used $253$ videos for training, $32$ for validation and $32$ for testing. 

\begin{table}[h]
\centering
\begin{tabular}{|l|c|c|}
\hline
\textbf{Split} & \textbf{PTSD samples} & \textbf{NO PTSD samples} \\ \hline
Training & 253 & 253 \\ \hline
Validation & 32 & 32 \\ \hline
Testing & 32 & 32 \\ \hline
\end{tabular}
\caption{Distribution of the classes in the splits.}
\label{table:single_split_dist}
\end{table}

\paragraph{N-fold cross validation}
To further evaluate the effectiveness of our baseline models and any new approach for video PTSD diagnosis, we used n-fold cross validation strategy with $n=3$ for the splitting of training and testing sets. By doing so, we ensure that our obtained results can be generalized to an independent subset of data. We also prevent overfitting and selection bias.

\subsection{Audio baseline model}
\label{subsec:audio_model}
Audio can be a powerful asset to identify emotions or affects from the speech \cite{SCHONEVELD20211}. More recently, several approaches have been proposed for mental disorders diagnosis based on audio data \cite{REJAIBI2022103107,MUZAMMEL2020100005,MUZAMMEL2021106433,othmani2021towards}. Audio signal has been also used in trauma prediction \cite{banerjee2019deep}. 

Therefore, we propose a baseline model for binary audio classification into With PTSD and Without PTSD.
The model used for audio classification is based on the VGGish model~\cite{hershey2017cnn}. Our actual implementation is based on the Keras implementation\footnote{https://github.com/DTaoo/VGGish} of the original VGGish model. The VGGish model is an adaption of VGG16 for audio files. Our starting point is the original VGGish model, pre-trained on the Audio Set dataset~\cite{gemmeke2017audio}.

The model architecture consists of four blocks of convolution layers. Each block is followed by a max-pooling layer with a kernel size of $2\times2$ and a stride of $2\times2$. The first block consists of a single convolutional layer which outputs 64 feature maps. The second block has a single convolutional layer with an output of 128 feature maps. The third block comprises two convolutional layers with 256 output feature maps each. The third block is also composed of two convolutional layers that both output 512 feature maps. All convolutional layers have a convolutional kernel of size $3\times3$ and a stride of $1\times1$.

\begin{figure*}[!t]
    \centering
    \includegraphics[width=1.0\textwidth]{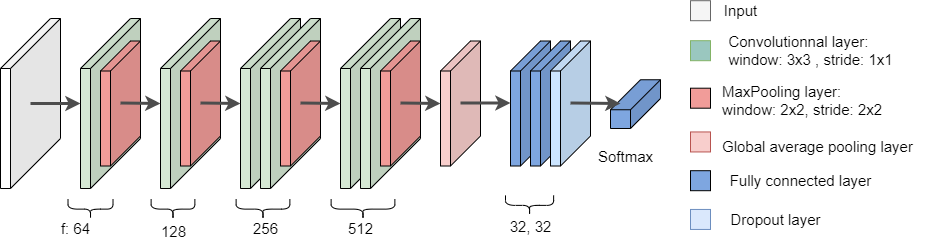}
    \caption{VGGish-based audio baseline model's architecture. The size of the feature maps (f) of each convolutional and fully-connected layer are shown below each block of
operations. }
    \label{fig:vggish}
\end{figure*}

On top of this, for the sake of our work, we replaced the original last three fully connected layers with a global pooling layer and two fully connected layers of $32$ units each. The number of units of these layers were found empirically by searching through a set of values. These layers are followed by a dropout layer and a fully connected layer of 2 units with a softmax activation representing the two classes. Fig.~\ref{fig:vggish} shows a representation of the modified VGGish model's architecture.

We pass to the model the log-mel-spectograms of each audio of size $2976\times64$. These inputs were computed using the original VGGish implementation\footnote{https://github.com/tensorflow/models/tree/master/research/audioset/vggish} in a window size of 30 seconds.

\subsection{Visual baseline model}
\label{subsec:visual_model}
The second proposed baseline model aims to predict PTSD based on the video frames. Video frames contain visual, spatial and temporal information to retrieve in order to extract PTSD patterns to discriminate between healthy and PTSD subjects. The proposed approach is based on PTSD symptoms recognition from patients’ responses to questions asked by an interviewer. 

In our proposed approach, the first step of the visual-based classification approach consists in pre-processing the data. Video Frames (images) are pre-processed in order to separate the patient’s face (interviewee) from that of the interviewer. Indeed, from each video in the dataset, only 60 frames containing face of the interviewee are extracted manually and are considered for training the visual baseline model.

Once interviewee (patient or control subject) frames are selected, the face is extracted using the MTCNN model~\cite{zhang2016joint}. Interviewee frames and face selection is an important step in the proposed approach and it impact considerably the performances of the proposed approach for visual PTSD classification.

For our visual baseline model, we used ResNet50v2~\cite{he2016identity}, originally trained on the ImageNet dataset~\cite{deng2009imagenet} for extracting meaningful features.
Then, the extracted features from the ResNet50v2 are fed to a sequence model consisting of recurrent layers of LSTM \footnote{https://keras.io/api/layers/recurrent\_layers/lstm/}.
Our hybrid architecture is popularly known as a CNN-RNN architecture.

ResNet50v2 is a modified version of ResNet50 that performs better due to a modification in the propagation formulation of the connections between blocks. Our actual implementation is based on the Keras implementation\footnote{ResNet50v2: https://keras.io/api/applications/resnet/\#resnet50v2-function}. The fully connected layer on the top of the model is removed for feature extraction. This technique of training a model from its pre-trained weights is called transfer learning, that focuses on storing knowledge gain, while solving one problem and applying it to a different but related problem. Note that transfer learning is a very useful technique when the size of dataset is small, as in our case. 

\begin{figure*}[!t]
    \centering
    \includegraphics[width=1.05\textwidth]{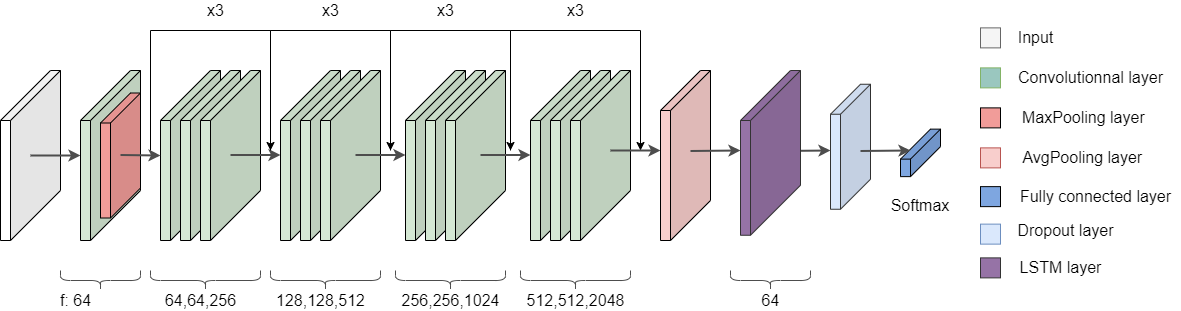}
    \caption{Resnet50v2 and LSTM-based visual baseline model's architecture. The size of the feature maps (f) of each convolutional and fully-connected layer are shown below each block of
operations.}
    \label{fig:resnet}
\end{figure*}

The resulting vector is passed through a fully connected layer for further classification as With PTSD or Without PTSD related videos. Fig.~\ref{fig:resnet} shows a representation of our proposed visual baseline model's architecture based on Resnet50v2 and LSTM.

 
\subsection{Text baseline model\label{subsec:text_model}}

Natural language processing (NLP) offers a valuable approach in recognizing the underlying thought in the text by understanding the meanings of individual words as well as the context which has been investigated and harnessed in several fields like sentiment analysis and emotion recognition~\cite{abualigah2020, desmet2013, batbaatar2019}. Thus, we consider it useful to utilize the NLP methods and tools in classifying the individuals by whether they have PTSD or not based on their transcriptions. Since we are working with videos, we need to : (1) extract the audio signal and (2) to convert it to transcribed text before (3) being fed to an automatic speech recognition (ASR) model. Our text classification network uses BERT~\cite{devlin2018} language model as the backbone to get the contextualized word embeddings which are then fed into a fully connected classification layer. We detail the ASR and text classification pipelines in the following sections.

\begin{figure*}[]
    \centering
    \includegraphics[width=1.0\textwidth]{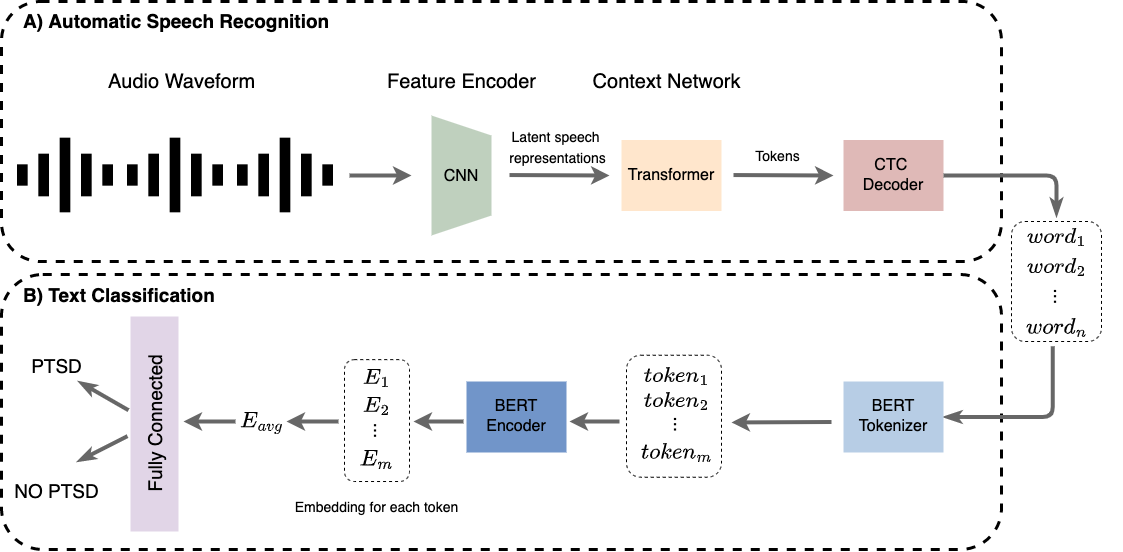}
    \caption{The pipeline of our text baseline model. \textbf{A) Automatic Speech Recognition:} To get the transcript from the audio waveform for each sample in our dataset, we used the wav2vec 2.0~\cite{baevski2020} architecture which is composed of a CNN feature extractor that feeds the latent speech representations into a Transformer to obtain the tokens which are then decoded to one of the 29 letters plus a space character using a CTC decoder. \textbf{B) Text Classification:} Given the plain text transcript of a sample, our aim is to classify it by PTSD or NO PTSD via the textual meaning. Thus, we first tokenize the text using the BERT tokenizer then we feed the tokens into the BERT encoder to get the embeddings for each token. We get an average text embedding for that sample by computing the mean of embeddings over the token dimension. Using one fully connected layer on top of BERT encoder handles the binary classification task.}
    \label{fig:text_baseline}
\end{figure*}

\subsubsection{Automatic Speech Recognition}
Classical ASR approaches rely on a large amount of transcript annotations of the speech audio to train, which is not feasible considering the human effort required. \\
We used the wav2vec 2.0~\cite{baevski2020} architecture as our speech-to-text converter which uses a self-supervised approach to train on a limited amount of labelled data. As shown in Fig.~\ref{fig:text_baseline}.A, the model is composed of three blocks : \textbf{i) Feature Encoder} is a multi-layer convolutional neural network that learns the latent audio representations for $T$ time steps from the raw audio. \textbf{ii) Context network} is a Transformer based architecture that takes the output of the feature encoder and generates $T$ representations carrying the information from the whole audio sequence. As in masked language modelling~\cite{devlin2018}, a ratio of the learned speech representations coming out of the feature encoder are randomly masked before feeding into the context network. To discretize the learned speech representations, \textbf{iii) a quantization module} is used by taking the output of the feature encoder with no masking as input and choosing a speech unit for each representation. The model jointly learns the quantized speech units and contextualized latent representations which boosts the performance significantly.

In the pre-training phase, the aforementioned blocks are trained in an end-to-end manned on the unlabeled data. Fine-tuning is done by minimizing the Connectionist Temporal Classification (CTC) loss~\cite{graves2006, baevski2019} on the labelled data. Nonetheless, we neither pre-trained nor the fine-tuned the ASR model on our dataset, \textit{i.e.} we used the ASR model just for inference, since we do not have the labels for the ASR task. We just executed inference on the audio files to get the transcribed text using a publicly available pre-trained and fine-tuned wav2vec 2.0 model. Also, we have long videos in our dataset with an average duration of 2 minutes, which makes it not applicable to directly run inference on the whole audio file because of the insufficient GPU memory. Thus, we needed to split the audio into small chunks. However, simply splitting the audio into a fixed duration of chunks may lead to poor results since we can cut in the middle of a spoken word. Although one possible solution is splitting based on the silence periods, it is not practical since we need to determine a threshold for loudness which is not an one-size-fits-all value. Hence, we run inference on partially overlapping chunks so that the model has the correct context focused in the center. The amount of overlapping is defined as stride. Then, we drop the logits on the sides, \textit{i.e.} take the logit at the center. Chaining the remaining logits gives us better results than the naive chunking.

\subsubsection{Text Classification}
Language models (LMs) are powerful NLP tools used in many areas like machine translation, question answering, named entity recognition and text classification due to their ability to understand the underlying abstract meaning of the natural language. However, training such data-hungry models require a large and diverse dataset to grasp the characteristics of the natural language. Trying to train a LM from scratch on a small corpus is most probably going to be not enough for it to generalize well.

BERT~\cite{devlin2018} model addresses this issue by proposing an architecture pre-trained on a large unlabelled text which can be used on the downstream tasks by fine-tuning with a single layer on top of it. BERT uses a deep bidirectional Transformer architecture to jointly train on two objectives: masked language modelling (MLM) and next sentence prediction. It utilizes a WordPiece tokenizer which allows BERT to recognize relevant words since they may share the same tokens. BERT model outputs a contextualized word embedding vector for each input token, \textit{i.e.} the representation for each token is built based on its surrounding tokens, thus the same words with the different meanings can be distinguishable. 

We used the BERT language model for text classification in our text baseline (Fig.~\ref{fig:text_baseline}.B). We tokenize the transcribed speech of each sample obtained from the ASR using the BERT tokenizer that outputs a sequence of tokens. Each sequence starts with \verb|[CLS]| and ends with \verb|[SEP]| tokens. We use \verb|[PAD]| tokens to pad the sequence to a constant length of 512. We truncate the longer sequences. Then, we use the base BERT encoder to get the contextualized embeddings for each token. We get the text embedding by averaging the embeddings over the token dimension. We feed the text embedding into a fully-connected layer that acts as a binary classifier.

\begin{table*}
\centering
\begin{tabular}{|p{0.250\linewidth}|p{0.45\linewidth}|c|}
\hline

\textbf{Hyperparameters} & \textbf{Set of considered values}                                  & \textbf{Optimal} \\ 
\hline \hline
Optimizer                & Adam, Adamax, Adagrad, Adadelt, SGD                  & \textbf{Adam}             \\ \hline
Learning rate            & 1e-05 to 1e-02                                        & \textbf{3e-05}            \\ \hline
Units                    & 32, 40, 48, 56, $\cdots$, 248, 256                      & \textbf{32}               \\ \hline
Dropout rate             & 0.1, 0.2, 0.3, 0.4, 0.5                              & \textbf{0.4}              \\ \hline
Kernel constraint        & 0.1, 0.2, 0.3, $\cdots$, 0.9, 1.0, 2.0, 3.0            & \textbf{3.0}              \\ \hline
Batch size               & 8, 16, 24, 32, $\cdots$, 120, 128                      & \textbf{128}              \\ \hline
Epochs                   & 1, 2, 3, 4, 5, 10, 15, 20, 30, 40, 50, $\cdots$, 100   & \textbf{100}              \\ \hline
\end{tabular}

\medskip
(a) Audio baseline model.
\medskip

\begin{tabular}{|p{0.250\linewidth}|p{0.45\linewidth}|c|}
\hline
\textbf{Hyperparameters} & \textbf{Set of considered values}                                          & \textbf{Optimal} \\ 
\hline
\hline
Optimizer                & Adam, Adamax, Adagrad, Adadelta, SGD, RMSprop, Nadam, Ftrl   & \textbf{SGD}             \\ \hline
Learning rate            & 3e-04, 1e-03, 1e-02, 1e-01                                       & \textbf{3e-04}            \\ \hline
LSTM units               & 2, 4, 8, 16, 32, 64                                          & \textbf{64}               \\ \hline
Dropout rate             & 0.2, 0.25, 0.3, 0.35, 0.4, 0.45, 0.5                         & \textbf{0.5}              \\ \hline
Momentum                 & 0.0, 0.1, 0.2, 0.9                                           & \textbf{0.0}              \\ \hline
\end{tabular}

\medskip
(b) Visual baseline model.

\caption{Hyperparameters search and optimization using KerasTuner for the audio and visual baseline models.}
\label{table:hyperparams}
\end{table*}

\section{Results on the baseline model}
\label{sec:results}
In this section, we present the results of the experiments performed using the baseline models on the PTSD in-the-wild dataset. In Section~\ref{subsec:metrics}, we describe the metrics used to evaluate the performances of our deep learning based baseline models. In Section~\ref{subsec:implementation}, we give some details on the implementation of our models, with a specific interest to the used parameters. Experimental results obtained with our audio baseline model, visual baseline model and text baseline model are described in Section~\ref{subsec:res_audio}, Section~\ref{subsec:res_visual} and Section~\ref{subsec:res_text} respectively.

\subsection{Evaluation metrics\label{subsec:metrics}}
We are interested in evaluating a binary classification task. There are two classes: With PTSD and Without PTSD. As we have to test the presence of a PTSD, we define the $TP$ (true positives), $FN$ (false negatives), $TN$ (true negatives) and $FP$ (false positives) values as such:
\begin{itemize}
    \item $TP$ is the number of samples labelled With PTSD and predicted as With PTSD,
    \item $FN$ is the number of samples labelled With PTSD and predicted as Without PTSD,
    \item $TN$ is the number of samples labelled Without PTSD and predicted as Without PTSD,
    \item $FP$ is the number of samples labelled Without PTSD and predicted as With PTSD.
\end{itemize}
From the confusion matrix, we compute the different popular classification metrics : the accuracy, the precision, the recall and the F1-score. The different metrics are described in the following. \\
\textbf{Accuracy} is the fraction of correctly classified samples and computed as:
\begin{equation}
    Accuracy = \frac{TP + TN}{TP + TN + FP + FN}
\end{equation}
\textbf{Precision} indicates the strength of the model to classify a negative sample as non-positive, which can be computed as:
\begin{equation}
    Precision = \frac{TP}{TP + FP} \\
\end{equation}
\textbf{Recall} indicates the ability of the model to find all the positive samples and can be computed as: 
\begin{equation}
    Recall = \frac{TP}{TP + FN} \\
\end{equation}
\textbf{F1-score} is the harmonic mean of precision and recall, thus can be computed as:
\begin{equation}
    F1-score = 2 \times\frac{Precision \times Recall}{Precision + Recall}
\end{equation}
or also:
\begin{equation}
    F1-score = \frac{2 \times TP}{2 \times TP + FP + FN}
\end{equation}

\subsection{Implementation details\label{subsec:implementation}}

Finding the best configuration of the hyperparameters of deep neural networks -- both related to the model and to the training -- is a computationally expensive task. In fact, the validation error function is very costly to evaluate for each hyperparameter value in order to find the best value by solving an optimization problem. This function is often not convex: it is then impractical to find its global minimum. 

\paragraph{Parameters setting of the audio baseline model}
To find the best hyperparameters, we used a bayesian optimizer built-in KerasTuner~\cite{omalley2019kerastuner}. KerasTuner is a tool for hyperparameters search and optimization. Bayesian optimization uses principles based on Bayes theorem to efficiently and effectively conduct a search of global optimization problem. We describe the list of hyperparameters and the sets of considered values in Table~\ref{table:hyperparams}.a. The optimal hyperparameters are presented in the last column. Then, we selected the Adam optimizer~\cite{kingma2014adam} with a learning rate of 3e-05. The model has a dropout rate of 0.4 to prevent overfitting after layers which have 32 units. The kernel constraint is equal to 3.0. We consider a batch size of 128 samples and the number of epochs is set to 100. 

We also had to prepare the audio files for the training. To train the model, we extracted audio files from the videos. The classification into With PTSD and Without PTSD through audio can be sensitive to the interference of the interviewer's speech and to background sound. For this reason, we dropped some audios of the two categories of type of trauma: Plane crash and Terrorist attack videos because the speech of the interviewer interfered a lot with the interviewee's speech and the background sound makes some audios not clearly audible. Moreover, the audios were split into chunks of 30 seconds. Thus, we dropped the chunks that do not contain the speech of the interviewee and those in which the interviewer speaks the most.

In addition, we expand the labeled audio training
sets by using data augmentation techniques. Three types of audio augmentation techniques are then performed on the audio frames to perturb the raw audio signals and generate new ones:
\begin{itemize}
    \item \textbf{Noise injection}: a random noise is added to the audio signal. The noise is multiplied by an \(\alpha\) factor;
    \item \textbf{Pitch change}: The pitch of the audio is changed by a factor 0.5, 2 and 3 in semi-tones;
    \item \textbf{Time shifting}: The samples are moved forward or backward by a given number of seconds.
\end{itemize}

\paragraph{Parameters setting of the visual baseline model}
We also used KerasTuner to find the best hyperparameters, but using the built-in Random Search algorithm as a tuner. The Random Search algorithm bases its optimization strategy on a stochastic process.  The list of hyperparameters and their sets of considered values are presented in Table~\ref{table:hyperparams}.b. The optimal hyperparameters can be found in the last column. Then, we selected the SGD (Stochastic Gradient Descent) optimizer with a learning rate of 3e-04 and a momentum of 0. The model has a dropout rate of 0.5 to prevent overfitting after LSTM layers and its 64 units. 

\paragraph{Parameters setting of the text baseline model}
For each video in the dataset, we extracted the audio and then run inference using the base wav2vec 2.0 model that is pre-trained and fine-tuned on the 960 hours of Librispeech speech audio (wav2vec2-base-960h). The chunk length is 60 seconds and the stride is 10 seconds. We have written each transcript to the disk. For the text classification, we used the pre-trained base uncased BERT model for English (bert-base-uncased) to create the contextualized representations. Just before the fully-connected layer, we added a dropout layer~\cite{srivastava2014} with a dropout rate of 0.1 to prevent the overfitting. We fine-tuned the BERT on the aforementioned text classification task for 5 epochs using the AdamW~\cite{loshchilov2017} optimizer with the learning rate of 5e-05, betas of (0.9, 0.999) and weight decay of 0.01. For single train/validation/test split methodology, we monitored the validation F1-score during the training and chose the model with the highest validation F1-score for testing. In N-Fold cross-validation, we trained for 3 folds as well and evaluated the model using the left-out testing set as usual.


\begin{table*}[]
\centering
\begin{tabular}{|l|c|c|c|c|}
\hline
\textbf{Baseline model} & \textbf{Accuracy} & \textbf{Precision} & \textbf{Recall} & \textbf{F1-score}  \\ \hline
\textbf{Audio} & 0.93 & 0.94 & 0.93 & 0.93 \\ \hline
\textbf{Visual} & 0.82 & 0.82 & 0.82 & 0.82 \\ \hline
\textbf{Text} & 0.98 & 1.00 & 0.97 & 0.98 \\ \hline
\end{tabular}
\caption{\label{table:test_results}Classification results on the testing set as a function of the used baseline model.}
\end{table*}

\begin{table*}[]
\centering
\begin{tabular}{cl|c|c|c|c|}
\cline{3-6}
\multicolumn{1}{l}{}                                                                                               &                  & \textbf{Accuracy} & \textbf{Precision} & \textbf{Recall} & \textbf{F1-score} \\ \hline
\multicolumn{1}{|c|}{\multirow{4}{*}{\textbf{\begin{tabular}[c]{@{}c@{}}Audio\\ baseline\\ model\end{tabular}}}}   & Fold 1           & 0.80              & 0.80               & 0.80            & 0.80              \\ \cline{2-6} 
\multicolumn{1}{|c|}{}                                                                                             & Fold 2           & 0.93              & 0.93               & 0.93            & 0.93              \\ \cline{2-6} 
\multicolumn{1}{|c|}{}                                                                                             & Fold 3           & 0.92              & 0.92               & 0.92            & 0.92              \\ \cline{2-6} 
\multicolumn{1}{|c|}{}                                                                                             & \textbf{Average} & \textbf{0.88}     & \textbf{0.88}      & \textbf{0.88}   & \textbf{0.88}     \\ \Xhline{2.5\arrayrulewidth}
\multicolumn{1}{|c|}{\multirow{4}{*}{\textbf{\begin{tabular}[c]{@{}c@{}}Visual \\ baseline\\ model\end{tabular}}}} & Fold 1           & 0.88              & 0.88               & 0.88            & 0.88              \\ \cline{2-6} 
\multicolumn{1}{|c|}{}                                                                                             & Fold 2           & 0.82              & 0.82               & 0.82            & 0.82              \\ \cline{2-6} 
\multicolumn{1}{|c|}{}                                                                                             & Fold 3           & 0.82              & 0.82               & 0.82            & 0.82              \\ \cline{2-6} 
\multicolumn{1}{|c|}{}                                                                                             & \textbf{Average} & \textbf{0.84}     & \textbf{0.84}      & \textbf{0.84}   & \textbf{0.84}     \\ \Xhline{2.5\arrayrulewidth}
\multicolumn{1}{|c|}{\multirow{4}{*}{\textbf{\begin{tabular}[c]{@{}c@{}}Text\\ baseline\\ model\end{tabular}}}}    & Fold 1           & 0.99              & 0.98               & 0.99            & 0.98              \\ \cline{2-6} 
\multicolumn{1}{|c|}{}                                                                                             & Fold 2           & 0.97              & 0.96               & 0.99            & 0.97              \\ \cline{2-6} 
\multicolumn{1}{|c|}{}                                                                                             & Fold 3           & 0.97              & 0.97               & 0.97            & 0.97              \\ \cline{2-6} 
\multicolumn{1}{|c|}{}                                                                                             & \textbf{Average} & \textbf{0.98}     & \textbf{0.97}      & \textbf{0.98}   & \textbf{0.98}     \\ \hline 
\end{tabular}

\caption{\label{table:cross_val_results}3-fold cross-validation classification results as a function of the used baseline model.}
\end{table*}

\subsection{Results obtained using the audio baseline model\label{subsec:res_audio}}
In order to evaluate how the model performs with the PTSD-in-the-wild dataset, we trained the model using the training, validation and testing sets for 100 epochs. 

The classification results on the unseen testing set are presented in Table~\ref{table:test_results} (first row). The accuracy is equal to 0.93. Moreover, a 0.93 F1-score value indicates that the trade-off between precision (equal to 0.94) and recall (equal to 0.93) is interesting. Thus, the audio baseline model achieves a good performances in the classification task, and it is very accurate in distinguishing between With PTSD and Without PTSD samples. 

In addition, we performed a 3-fold cross-validation to validate the results. The results are very close to the previous results. Table~\ref{table:cross_val_results} summarizes the results of the cross-validation. The model has an average accuracy of 0.88, with a standard deviation of 0.05. It performed better on the second and the third folds, with results similar to those obtained using the testing set (as presented in Table~\ref{table:test_results}). The overall results show that the audio baseline model is able to learn and to generalize from a new dataset.

\subsection{Results obtained using the visual baseline model\label{subsec:res_visual}}
The classification results using the visual baseline model on the testing set are presented in Table~\ref{table:test_results} (second row). With a value of 0.82, we can see that the accuracy is slightly worse than the value obtained using the audio baseline model. It is still interesting, as well as the value of the F1-score: they both remain high, which shows good classification performances.

Moreover, the 3-fold cross validation results are provided in Table~\ref{table:cross_val_results}. The visual baseline model has an average accuracy of 0.84 and performed better on the first fold. Nevertheless, whatever the considered fold, the results are similar to those obtained using the testing set (as presented in Table~\ref{table:test_results}). These results prove the generalization ability of the visual baseline model.





\subsection{Results obtained using the text baseline model\label{subsec:res_text}}
In Table~\ref{table:test_results} (third row), we report the classification results obtained using the text baseline model on the testing set. Being equal to 0.98, the accuracy is very close to the maximum value. Moreover, the F1-score is equal to 0.98. It indicates that the trade-off between precision (equal to 1.00) and recall (equal to 0.97) is excellent.

To further see the generalizability of the text baseline model, we also reported the 3-fold cross validation results in Table~\ref{table:cross_val_results}. The text baseline model has an average accuracy of 0.98. The performances are almost the same on every fold and comparable to those using the testing set (as presented in Table~\ref{table:test_results}): the classification is nearly perfect. Therefore, the text baseline model outperforms the two other baseline models using audio or visual information to distinguish between With PTSD and Without PTSD samples. The text is the best modality among the three evaluated modalities in PTSD recognition and detection.

\section{Discussion and conclusion}
\label{sec:conclusion}
The analysis of human mental health and disorders is a very complex, challenging and sensitive problem. 
The lack of understanding of mental illness and the prejudice and the negative beliefs towards it have caused universally the stigma of mental illness. Consequently, a more attention is required from the scientific community to join forces to improve mental health and to propose innovative solutions for better and more precise diagnosis, prognosis and assistance of the patients. The development of innovative clinical applications for mental disorders diagnosis is plagued by ethical and privacy issues. However, artificial intelligence (AI) research for psychiatry and mental health faces several strict limitations : (1) the lack and the small size of the available datasets. (2) the limited available data and modalities and (3) the absence of audiovisual data that could be relevant for the diagnosis. To overcome these problems, AI can be used as a tool to guarantee the privacy and the protection of sensitive data without being a threat \cite{kusters2020interdisciplinary}.

Artificial intelligence and affective computing based tools and systems can play a key role in the diagnosis of several mental disorders, the follow-up of mental state of patients and the development of personalized medicine in psychiatry. One of the main limitations is the lack of datasets and cohorts. In this paper, we proposed and we made publicly available for academic research a new dataset called PTSD-in-the-wild dataset for trauma and post-traumatic stress disorder recognition and analysis from video data. The dataset is collected in unconstrained environments and conditions. It means that the video are acquired
in real-world scenarios without any lab-controlled conditions. To be able to cope with such challenging conditions, the proposed approaches should be robust to indoor, outdoor, different color backgrounds, occlusions, background clutter and face misalignment. The proposed dataset can be used for binary learning and classification of PTSD. However, it can not be used for PTSD assessment as no self-assessment questionnaire like the Post-traumatic Stress Disorder Checklist (PCL) test is performed. The PCL-5 is a 20-item, widely used DSM-correspondent self-report that assesses the 20 DSM-5 symptoms of PTSD.

In this paper, we propose a baseline models for PTSD recognition using the PTSD-in-the-wild dataset. Two benchmarks are given for the evaluation and the comparison of proposed approaches on the PTSD-in-the-wild dataset. The visual, audio and textual baseline models show very promising and high performing results. However, our baseline visual model is very dependant on the manual frame selection of the interviewee. The performance of the audio baseline model strongly dependant on the drop of the two categories of plane crash and Terror attack videos because of the strong intervention of the interviewer during the interview and the bad quality of the video and the high ratio signal to noise. The three modalities of facial images, audio signals and speech are very accurate to recognize and to discriminate PTSD patterns during interviews sessions of healthy control subjects and PTSD subjects who experienced a traumatic event. The high performances of the baseline models can be explained by the big difference between the content of the two classes: (1) the first class of interviews of individuals who experienced a traumatic event and who have PTSD symptoms and talking about the traumatic event and showing severe emotional distress or physical reactions when reminding it and (2) the second class of interviews of random topics with actors and public figures, showing a positive affects, thoughts and attitudes.  
The question arises: Is it possible to discriminate between a PTSD person and a person with a negative affect (e.g. a sad person) or a person with anxiety disorder or also a depressed person ?

In future work, we are planned: (I) to propose an automatic approach to perform the frame selection and interviewee face detection and interviewee audio signal extraction to make our approach fully automatic and (II) to propose a second version of the PTSD-in-the-wild dataset with other low mental mood and affects conditions and disorders.

\ifCLASSOPTIONcaptionsoff
  \newpage
\fi



%



\bibliography{citation.bib}
\bibliographystyle{ieeetr}

%

\begin{IEEEbiography}[{\includegraphics[width=1in,height=1.25in,clip,keepaspectratio]{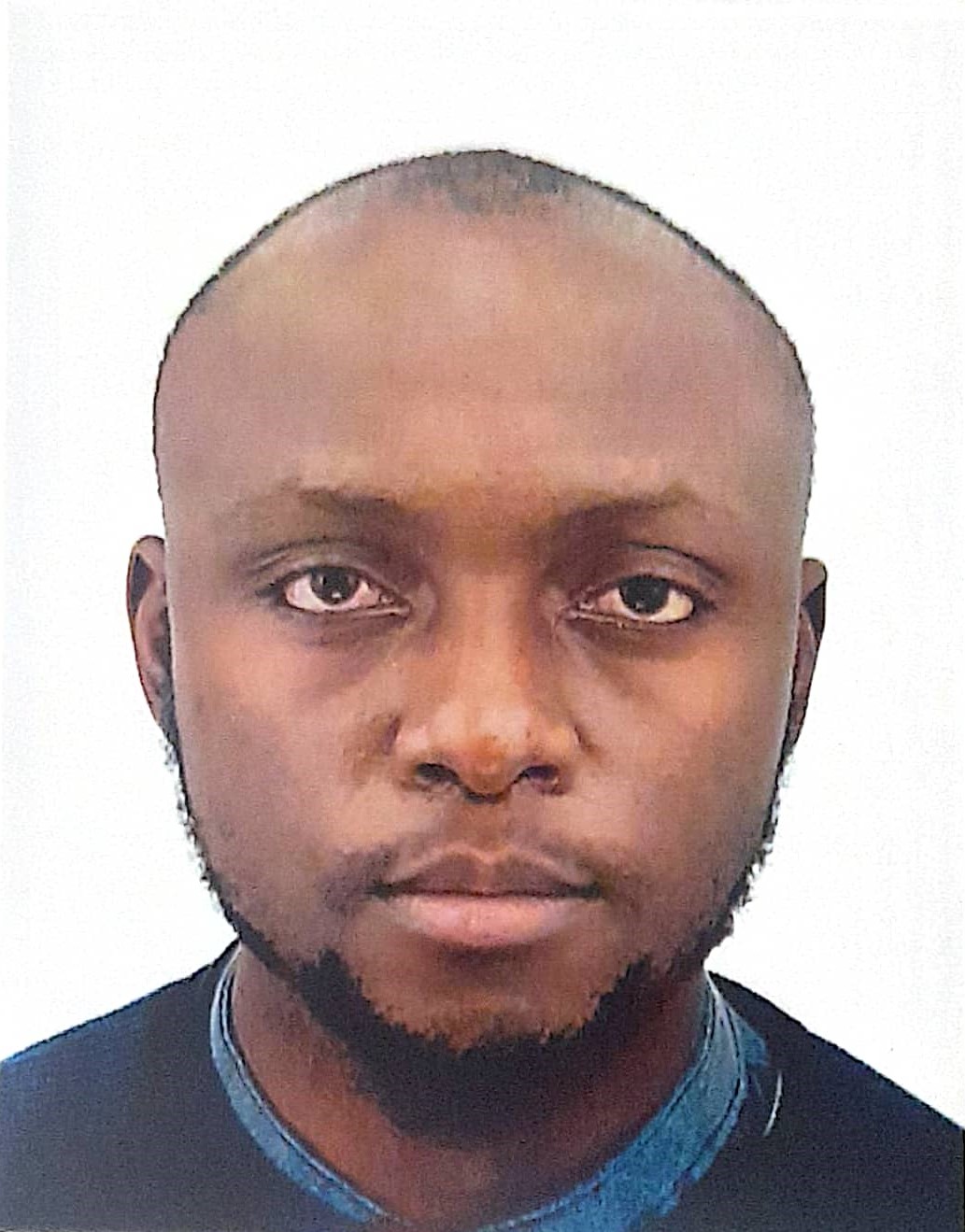}}]{Moctar Abdoul Latif Sawadogo}
is currently a Master student at University Paris City in France. He was a master intern at Laboratory of Images, Signals and Smart Systems of University Paris-Est Créteil (UPEC).
His main interests are machine learning and deep learning.
\end{IEEEbiography}

\begin{IEEEbiography}[{\includegraphics[width=1in,height=1.25in,clip,keepaspectratio]{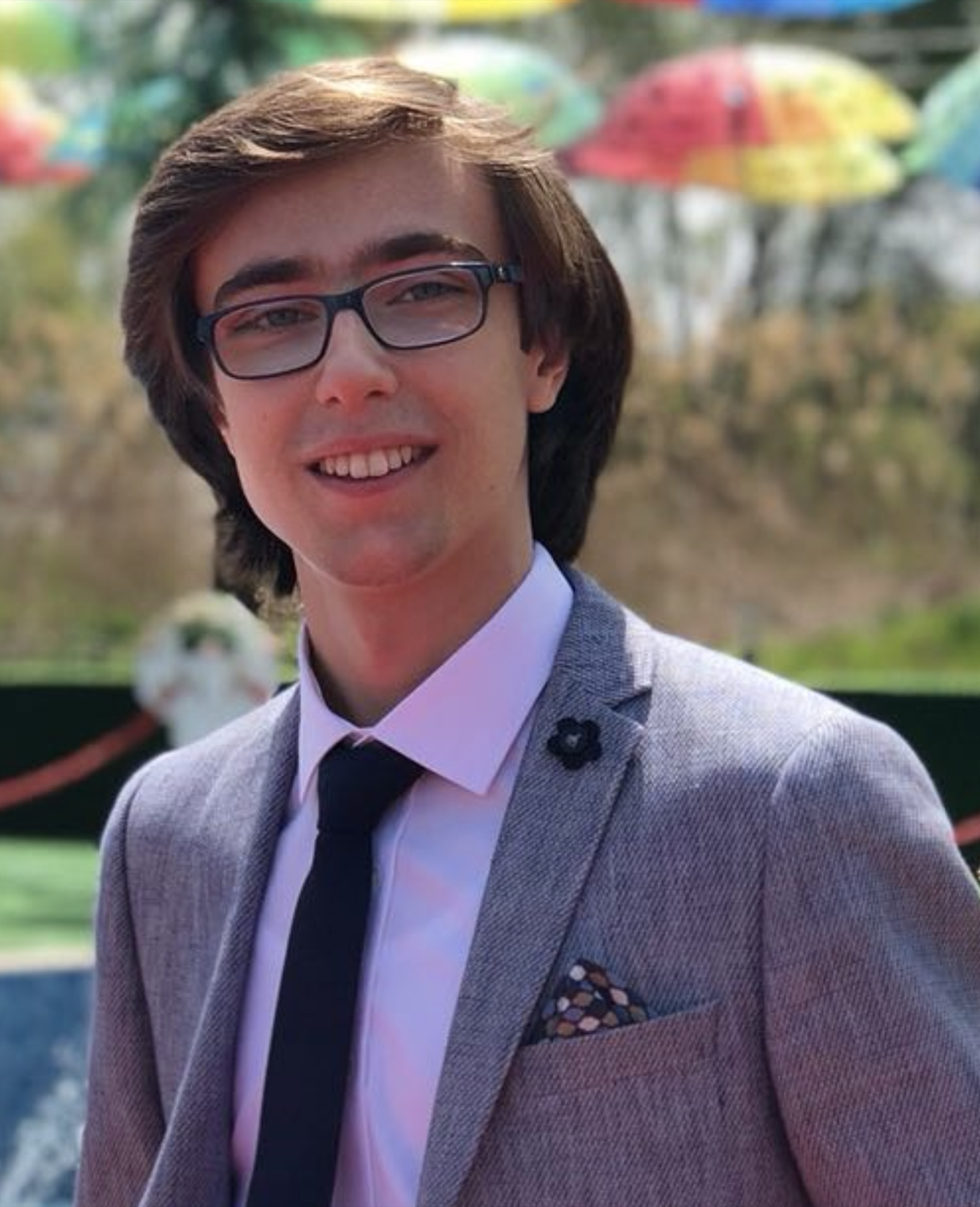}}]{Furkan Pala}
graduated from the Istanbul Technical University (ITU) Computer Engineering bachelor program in 2022. He is currently pursuing a double major degree in Mathematics Engineering at ITU. He published two conference papers about medical imaging accepted in MICCAI 2021 and 2022 during his internship at Brain And SIgnal Research \& Analysis (BASIRA) laboratory, ITU. His work focuses on deep learning, computer vision, medical imaging and NLP.
\end{IEEEbiography}

\begin{IEEEbiography}[{\includegraphics[width=1in,height=1.25in,clip,keepaspectratio]{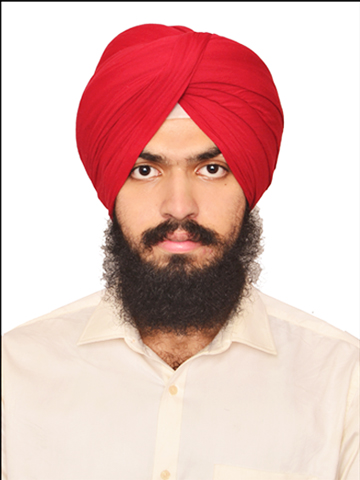}}]{Gurkirat Singh}
is currently pursuing Bachelor in Technology at National Institute of Technology, Warangal, India majoring in Computer Science. His main research interests include but are not limited to Robotics, Deep Learning and Computer Vision. 
\end{IEEEbiography}


\begin{IEEEbiography}[{\includegraphics[width=1in,height=1.25in,clip,keepaspectratio]{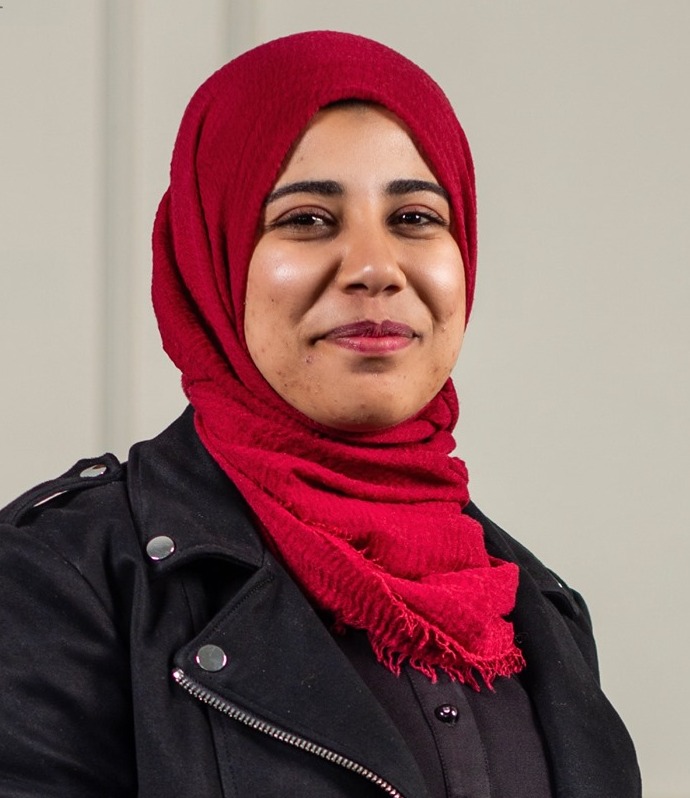}}]{Imen Selmi}
 is a computer science and applied Mathematics engineering student at The National Engineering School of Sfax (ENIS), University of Sfax. She worked on many projects in deep learning and computer vision and she was a master intern at Laboratory of Images, Signals and Smart Systems of University Paris-Est Créteil (UPEC).
\end{IEEEbiography}

\begin{IEEEbiography}[{\includegraphics[width=1in,height=1.25in,clip,keepaspectratio]{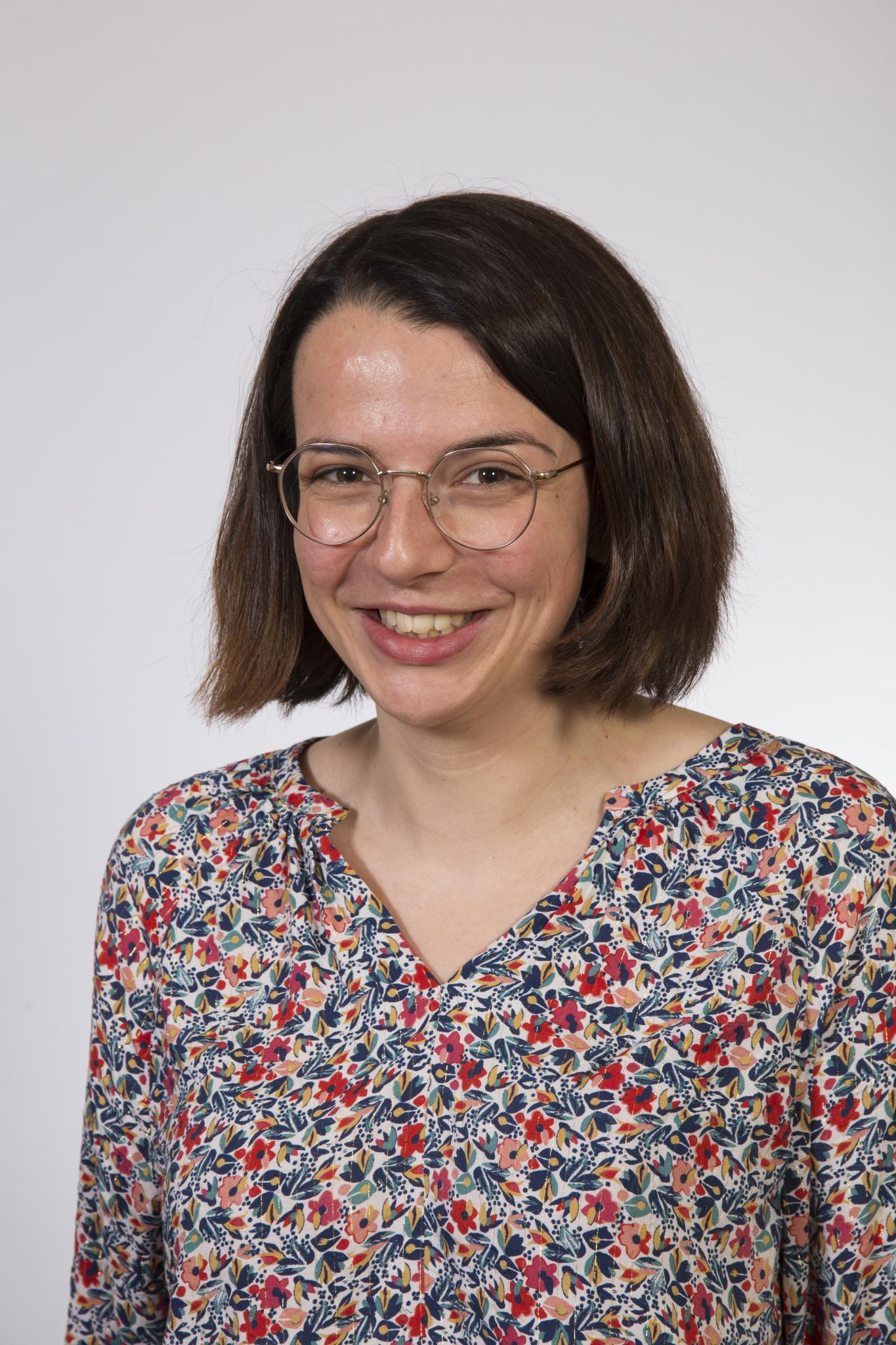}}]{Pauline Puteaux} received her M.S. degree in Computer Science and Applied Mathematics with specialization in Cybersecurity from the University of Grenoble, France, in 2017 and her PhD degree in computer science from the Université de Montpellier, France, in 2020. She is currently working as a researcher for the CNRS (French National Centre for Scientific Research) with the Centre de Recherche en Informatique, Signal et Automatique de Lille (CRIStAL), France. Her work has focused on multimedia security, and in particular, image analysis and processing in the encrypted domain. Since 2016, she has published eight journal articles and thirteen conference papers. She is a reviewer for Signal Processing (Elsevier), the Journal of Visual Communication and Image Representation (Elsevier), the IEEE Transactions on Circuits \& Systems for Video Technology, and the IEEE Transactions on Dependable and Secure Computing. Photo credit: © Xavier PIERRE / CNRS
\end{IEEEbiography}

\begin{IEEEbiography}[{\includegraphics[width=1in,height=1.25in,clip,keepaspectratio]{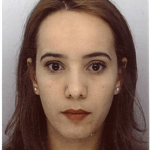}}]{Alice Othmani} is an associate professor at Université Paris-Est Créteil since 2017. Her research works concern developing computer vision and Artificial Intelligence solutions for Healthcare, emotional intelligence and psychiatry. She has been working in several international institutions like Ecole Normale Supérieure de Paris, Collège de France and Agency for Science, Technology and Research (A*STAR) in Singapore. 
\end{IEEEbiography}




\end{document}

%% file: graphics/trauma_distribution.tex
\definecolor{azure(colorwheel)}{rgb}{0.0, 0.5, 1.0}
\definecolor{bittersweet}{rgb}{1.0, 0.44, 0.37}
\definecolor{chromeyellow}{rgb}{1.0, 0.65, 0.0}
\definecolor{fluorescentorange}{rgb}{1.0, 0.75, 0.0}
\definecolor{deepsaffron}{rgb}{1.0, 0.6, 0.2}

\tikzset{
     lines/.style={draw=none},
}

\begin{tikzpicture}
    \pie[
        radius = 5.5,
        rotate = 180,
        style = {lines},
        scale font
    ]{
        68.45/War veteran,
        11.99/Plane crash,
        11.67/Sexual assault,
        3.47/Terrorist attack,
        0.32/Domestic abuse,
        1.27/Car accident,
        2.84/Other
    }
\end{tikzpicture}

%% file: graphics/gender_trauma_distribution.tex
\begin{tikzpicture}
\begin{axis}[ybar,
        legend pos=outer north east,
        legend cell align=left,
        bar width = 20pt,
        width = 7.5cm,
        xticklabels = {Male, Female},
        xtick=data,
        enlargelimits=0.25,
        x tick label style={
    		/pgf/number format/1000 sep=},
    ]

    \addplot coordinates {(0,226) (1,131)};
    \addplot coordinates {(0, 241) (1,76)};
    
    \legend {With PTSD, Without PTSD};
    
\end{axis}
\end{tikzpicture}